\documentclass[prl,aps,twocolumn,preprintnumbers,showpacs]{revtex4}

\begin{document}
\title{General formalism of Hamiltonians for realizing a prescribed evolution of a qubit}
\author{D. M. Tong $^{1,2}$,   Jing-Ling Chen $^1$, L. C. Kwek$^{1,3}$, C. H. Lai$^1$, and
C. H. Oh$^1$\footnote{Electronic address: phyohch@nus.edu.sg}}
\address{$^1$ Department of Physics, National University of Singapore, 10 Kent Ridge Crescent, Singapore 119260 \\$^2$ Department of Physics, Shandong Normal University, Jinan 250014, P R China\\ $^3$ National Institute of Education, Nanyang Technological University, 1 Nanyang Walk, Singapore 639798 }
\date{\today}
\begin{abstract}
We investigate the inverse problem concerning the
evolution of a qubit system, specifically we consider how one
can establish the Hamiltonians that account for the evolution
of a qubit along a prescribed path in the projected Hilbert space.
For a given path, there are infinite Hamiltonians which can
realize the same evolution. A general form of the Hamiltonians is
constructed in which one may select the desired one for
implementing a prescribed evolution. This scheme can be generalized
to higher dimensional systems.
\end{abstract}
\maketitle

\section{Introduction}
Evolution of a quantum system is completely determined by
the given dynamics. For a closed system, the density matrix
$\rho(t)$ of the system satisfies the evolution equation,
\begin{eqnarray}
\rho(t)=U(t)\rho(0)U(t)^\dagger , \label{dotrhot}
\end{eqnarray}
where  $U(t)=T\exp[-i \int_0^t H(t)dt]$ is the unitary operator
corresponding to the Hamiltonian $H(t)$. For an open system, its
evolution is generally nonunitary and therefore it cannot be
described by the above equation. However, an open system, denoted
as $S_a$, can always be regarded as a part of a larger closed
system, denoted as $S_{ab}$, comprising 
$S_a$ and its environment, denoted as $S_b$. The evolution of
combined system $S_{ab}$ now obeys Eq. (\ref{dotrhot}) whereas
the evolution of the open system $S_a$ is described by the Kraus
representation \cite{Kraus}
\begin{eqnarray}
\rho(t)=\sum\limits_{\mu} M_\mu\rho(0){M_\mu} ^\dagger ,
\label{rhokraus}
\end{eqnarray}
where Kraus operators $M_{\mu}$ satisfy $\sum\limits_\mu
M_\mu^\dagger M_\mu=I$. The Kraus operators are usually expressed
as $M_\mu=\langle \mu |U_{ab}(t)|0\rangle$, where
$U_{ab}(t)=T\exp[-i \int_0^t H_{ab}(t)dt]$ is determined by the
Hamiltonian $H_{ab}(t)$ of the combined system and
$|\mu\rangle $ are the orthogonal bases of the environment.
Clearly, given the Hamiltonian $H(t)$ for the closed system or the
Hamiltonian $H_{ab}(t)$ for the combined system, the evolution of
the density matrix $\rho(t)$ is completely determined through Eq.
(\ref{dotrhot}) or Eq. (\ref{rhokraus}).

It is interesting to study the inverse problem. Supposing the
density matrix $\rho(t)$ of a quantum system, be it closed or
open, is given as a time-dependent function, how can one obtain
the Hamiltonian? This is an interesting and nontrivial
issue because oftentimes in experiment, one needs to prepare 
a quantum system which is expected to evolve along a given path
on the surface or inside the Bloch sphere.
Certainly, one can always find this Hamiltonian through trials and errors.
To this end, one may look for some trial Hermitian
operators and require them to satisfy the
evolution equation with the given $\rho(t)$, however, this is
not easy. Indeed, it may happen that even if the form of the Hamiltonian
is eventually found, it may still be too difficult to set it up
experimentally. A suitable Hamiltonian must not only serve the
given evolution mathematically, it should also be realizable physically.
In particular, the problem becomes very difficult for nonunitary
evolutions of open systems albeit its solvability for unitary evolution
of closed systems.

The purpose of this paper is to put forward a general approach for
finding the appropriate Hamiltonian that
determines a given evolution of the systems. We restrict our
discussion to two level system (the qubit) which is generally the most
prevalent system used in quantum computation
and quantum information. We will provide a general formalism of the
Hamiltonians for realizing
an arbitrary prescribed evolution of the qubit system. The Hamiltonians are not unique. Our formalism gives a set of the equivalent
Hamiltonians. Both unitary and nonunitary evolutions are investigated respectively.

\section{Hamiltonians for realizing unitary evolution of a qubit}
For a qubit system, the density matrix can be expressed as
\begin{eqnarray}
\rho(t)=\frac{1}{2}(1+{\bf{r}} {\bf{\cdot \sigma}}),
\label{rhot0}
\end{eqnarray}
where $\bf{r}$, the Bloch vector, is a time-dependent vector function
with $|\bf{r}|\le 1$ and ${\bf{\sigma}}$ are the Pauli matrices.
When we refer to a given evolution,
it means that $\rho(t)$, as a matrix function of $t$, is explicitly given.
In other words, the movement of the Bloch vector $\bf{r}$ in Bloch sphere is known.
The most general form of $\rho(t)$ can be written as
\begin{eqnarray}
\rho(t)=\frac{1}{2} \left(\begin{array}{cc}
1+r\cos\theta&r\sin\theta~e^{-i\phi}
\\r\sin\theta~e^{i\phi}&1-r\cos\theta
\end{array}\right),
\label{rhot}
\end{eqnarray}
where $r=r(t)$,~$\theta=\theta(t)$ and $\phi =\phi (t)$
are arbitrary time-dependent real functions with
$0\leq r \leq 1,~0\leq \theta \leq \pi,~0\leq \phi \leq 2\pi$.
At $t=0$, we denote $r(0),~\theta (0)$ and $\phi(0)$ as $r_0,~\theta_0$ and $\phi_0$
respectively, and the general initial state reads
\begin{eqnarray}
\rho(0)=\frac{1}{2} \left(\begin{array}{cc}
1+r_0\cos\theta_0&r_0\sin\theta_0~e^{-i\phi_0}
\\r_0\sin\theta_0~e^{i\phi_0}&1-r_0\cos\theta_0
\end{array}\right).
\label{rho0}
\end{eqnarray}
Our aim is to find the general form of Hamiltonians that realize the evolution
defined by Eq. (\ref{rhot}) with the initial state (\ref{rho0}).
For unitary evolution, $r\equiv r_0$ is constant and there exist unitary operators $U(t)$
satisfying
\begin{eqnarray}
\rho(t)=U(t)\rho(0)U(t)^\dagger  . \label{r=uru}
\end{eqnarray}
If we can find the operators $U(t)$, the Hamiltonians $H(t)$ are easy to be obtained as
\begin{eqnarray}
H(t)&=&i\dot U(t)U(t)^\dagger  . \label{h}
\end{eqnarray}
Hence the problem reduces to the question of how to
find the unitary operators $U(t)$ satisfying
Eq. (\ref{r=uru}).  As mentioned before, the corresponding operators are not unique.
There are infinitely many unitary
operators $U(t)$ that can realize the same evolution, all of which are equivalent for
$\rho(t)$. We can find all the different but
equivalent Hamiltonians by finding all the $U(t)$. To this end, we first construct
one of the unitary operators. We find that the following operator
\begin{eqnarray}
\tilde{U}(t)= \left(\begin{array}{cc}
\cos\frac{\theta-\theta_0}{2}e^{-i\frac{\phi-\phi_0}{2}}
&-\sin\frac{\theta-\theta_0}{2}e^{-i\frac{\phi+\phi_0}{2}}
\\ \sin\frac{\theta-\theta_0}{2}e^{i\frac{\phi+\phi_0}{2}}&
\cos\frac{\theta-\theta_0}{2}e^{i\frac{\phi-\phi_0}{2}}\end{array}\right)
\label{tu}
\end{eqnarray}
satisfies Eq. (\ref{r=uru}) for an arbitrary evolution defined by
Eqs. (\ref{rhot}) and (\ref{rho0}). With the operator $\tilde{U}(t)$,
all the equivalent unitary operators can be constructed by
\begin{eqnarray}
U(t)=\tilde{U}(t)V(t),
\label{u=uv}
\end{eqnarray}
where $V(t)\in U(2)$ and satisfies the following commutation relation,
\begin{eqnarray}
[V(t)~,~\rho(0)]=0.
\label{vr}
\end{eqnarray}
To provide an explicit form for matrix $V(t)$,
we rewrite the initial density matrix $\rho(0)$
in the form of the orthogonal decompositions,
\begin{eqnarray}
\rho(0)=\sum\limits_{k=1,2}w_k\rho_k(0),
\label{r=wr}
\end{eqnarray}
where $w_1=(1+r_0)/2,~w_2=(1-r_0)/2$ and
\begin{eqnarray}
\rho_k(0)=\frac{1}{2} \left(\begin{array}{cc}
1+\delta_k\cos\theta_0&\delta_k\sin\theta_0~e^{-i\phi_0}
\\\delta_k\sin\theta_0~e^{i\phi_0}&1-\delta_k\cos\theta_0
\end{array}\right),
\label{rho012}
\end{eqnarray}
with $\delta_1=1,~\delta_2=-1$.
The above orthogonal decomposition is unique if $r_0\neq 0$ (we need not consider $r_0=0$
because in this case the Bloch vector does not move at all during the unitary evolution).
Thus, the unitary matrix $V(t)$ satisfying Eq. (\ref{vr}) can be written as
\begin{eqnarray}
V(t)=\sum\limits_{k=1,2}e^{i\alpha_k}\rho_k(0),
\label{v}
\end{eqnarray}
where $\alpha_k=\alpha_k(t)~~(k=1,2)$~ are arbitrary real parameters with
$\alpha_1 \mid_{t=0}=\alpha_2 \mid_{t=0}=0$.
From Eqs. (\ref{u=uv}) and (\ref{v}), the general form of the unitary operators
$U(t)$ is obtained as
\begin{eqnarray}
U(t)=\tilde{U}(t)\sum\limits_{k=1}^{2}e^{i\alpha_k}\rho_k(0).
\label{u=ur}
\end{eqnarray}
The corresponding general form of the Hamiltonians $H(t)$ is
\begin{eqnarray}
H(t)&=&i\dot{\tilde{U}}(t)\tilde{U}(t)^\dagger
-\sum\limits_{k=1}^{2}\dot \alpha_k
\tilde{U}(t)\rho_k(0)\tilde{U}(t)^\dagger \nonumber \\ & &
=\tilde{H}(t) -\sum\limits_{k=1}^{2}\dot \alpha_k \rho_k(t),
\label{h=iuu}
\end{eqnarray}
where $\tilde{H}(t)=i\dot{ \tilde{U}}(t)\tilde{U}(t)^\dagger $
and $\rho_k(t)=\tilde{U}(t)\rho_k(0)\tilde{U}(t)^\dagger $.
More explicitly, the Hamiltonians read
\begin{widetext}
\begin{eqnarray}
\hat{H}(t)=\frac{1}{2} \left(\begin{array}{cc}
\dot \phi -\dot \alpha_1(1+ \cos\theta) -\dot \alpha_2 (1- \cos\theta)&
(-i\dot \theta-\dot \alpha_1\sin\theta+\dot \alpha_2 \sin\theta)e^{-i\phi}
\\ (i\dot \theta-\dot \alpha_1\sin\theta+\dot \alpha_2 \sin\theta)e^{i\phi} &
-\dot \phi -\dot \alpha_1(1- \cos\theta) -\dot \alpha_2 (1+ \cos\theta)\end{array}\right).
\label{hm}
\end{eqnarray}\end{widetext}
The formula expressed by Eq. (\ref{hm}) provides the recipe for finding
all the Hamiltonians that can realize
the prescribed evolution defined by Eq. (\ref{rhot}) and (\ref{rho0}).
Since the evolution given by Eqs. (\ref{rhot}) and (\ref{rho0}) is arbitrary,
the Hamiltonians given in the above equation are the most general ones.
When a prescribed evolution $\rho(t)$ is given, {\sl{i.e.}}, $r=r_0$ is a known constant, 
 and $\theta(t)$, $\phi(t)$ are two known time dependent functions,
one can easily write down the corresponding Hamiltonians by directly substituting
the variables into Eq. (\ref{hm}).
Noting that there are two time-dependent parameters $\alpha_k(t)$ $(k=1,~2)$, which can
be arbitrary real functions of $t$ with $\alpha_k(0)=0$, one can establish the suitable
Hamiltonian for any special purpose by properly choosing the forms of $\alpha_k(t)$.

\section{Hamiltonians for realizing nonunitary evolution of a qubit}
We now turn to the nonunitary evolution of an open qubit  system.
For nonunitary evolution, the path traced out by the Bloch vector is a curve
inside the Bloch sphere with variable $r$.
In this case, Eq. (\ref{dotrhot}) is no longer valid.
To realize a prescribed evolution of the open system, we need to consider a larger closed
system, comprising the open system and an ancillary system which can also be
regarded as the
environment of the open system. For simplicity, we take the ancillary system to be
a qubit. We denote the qubit under prepared as $a$ and the ancillary qubit as
$b$, respectively. The two qubits constitute a combined system $S_{ab}$.
The combined system is a closed one, and its evolution is described by unitary
operator $U_{ab}(t)$.  The most general evolution of an open qubit system can still be
described in terms of expressions (\ref{rhot}) and (\ref{rho0}) with variable $r$.
Suppose the ancillary qubit is initially in the state $|0\rangle\langle 0|$,
the initial state of the combined system can be written as
$\varrho(0)=\rho(0)\otimes |0\rangle \langle 0|$, where $\rho(0)$ is an arbitrary given
initial state of the open system. Then, at time $t$, the density matrix $\varrho(t)$ of
the combined system reads
\begin{eqnarray}
\varrho(t)=U_{ab}(t)\rho(0)\otimes |0\rangle \langle
0|U_{ab}(t)^\dagger . \label{varrhot}
\end{eqnarray}
The reduced density matrix of the  open system is given by tracing out the state of the second qubit, $\it{i.e.}$
\begin{eqnarray}
\rho(t)={\rm tr}_b\varrho(t)={\rm tr}_b\{U_{ab}(t)\rho(0)\otimes
|0\rangle \langle 0|U_{ab}(t)^\dagger \}. \label{rhott}
\end{eqnarray}
The problem of preparing an open qubit system for a given evolution
reduces to that of finding the unitary operator $U_{ab}(t)$ such that the reduced matrix $\rho(t)$
remains the same as that assigned by Eqs. (\ref{rhot}) and (\ref{rho0}).
To this end, we rewrite Eq. (\ref{rhott}) by using the Kraus representation
\begin{eqnarray}
&&\rho(t)=\sum\limits_{\mu=0,1} M_\mu\rho(0){M_\mu} ^\dagger
,\nonumber\\&& \sum\limits_\mu M_\mu^\dagger M_\mu=I,
\label{rhotm}
\end{eqnarray}
where the Kraus operators $M_\mu=\langle \mu |U_{ab}|0\rangle$ ,
and $|\mu \rangle~(\mu=0,1)$ are the orthogonal bases of the ancillary qubit.
If we know the Kraus operators, we may inversely deduce the unitary
operators $U_{ab}(t)$. One readily verifies that the following operators
$\tilde{M}_\mu $ satisfy Eq. (\ref{rhotm}),
\begin{widetext}
\begin{eqnarray}
&\tilde{M}_0&= \left(\begin{array}{cc}
-\cos\frac{\theta}{2}\sin\frac{\theta_0}{2}-\sqrt{\frac{1-r}{1+r_0}}\sin\frac{\theta}{2}\cos\frac{\theta_0}{2}e^{i(\phi_0-\phi)}~~&\cos\frac{\theta}{2}\cos\frac{\theta_0}{2}e^{-i\phi_0}-\sqrt{\frac{1-r}{1+r_0}}\sin\frac{\theta}{2}\sin\frac{\theta_0}{2}e^{-i\phi}
\\-\sin\frac{\theta}{2}\sin\frac{\theta_0}{2}e^{i\phi}+\sqrt{\frac{1-r}{1+r_0}}\cos\frac{\theta}{2}\cos\frac{\theta_0}{2}e^{i\phi_0}~~&\sin\frac{\theta}{2}\cos\frac{\theta_0}{2}e^{i(\phi-\phi_0)}+\sqrt{\frac{1-r}{1+r_0}}\cos\frac{\theta}{2}\sin\frac{\theta_0}{2}
\end{array}\right),\nonumber \\
&\tilde{M}_1&= \sqrt{\frac{r+r_0}{1+r_0}}\left(\begin{array}{cc}
\cos\frac{\theta}{2}\cos\frac{\theta_0}{2}e^{i\phi_0}~~&\cos\frac{\theta}{2}\sin\frac{\theta_0}{2}\\\sin\frac{\theta}{2}\cos\frac{\theta_0}{2}e^{i(\phi+\phi_0)}~~&\sin\frac{\theta}{2}\sin\frac{\theta_0}{2}e^{i\phi}
\end{array}\right).
\label{mm}
\end{eqnarray}\end{widetext}
The choice of the Kraus operators is not unique. There are infinitely many (equivalent)
operators $M_\mu$ satisfying Eq. (\ref{rhotm})  for given $\rho(t)$ and $\rho(0)$.
$\tilde{M}_\mu$ is only one special choice. Correspondingly, there are also infinitely many
choices of $U_{ab}(t)$ satisfying Eq. (\ref{rhott}), so do $H(t)$.
With the known Kraus
operators $\tilde{M}_\mu(t)$, we can first construct one of the operator $U_{ab}(t)$,
and denote it as $\tilde{U}_{ab}(t)$.
In fact, $\tilde{U}_{ab}(t)$ is a $4\times 4$ unitary matrix, where the elements in the
first and the third columns are completely determined by the Kraus operators
$\tilde{M}_{\mu}$ with ${\tilde{U}}_{11}=({\tilde M}_0)_{11},
~{\tilde{U}}_{21}=({\tilde M}_1)_{11},
~{\tilde{U}}_{31}=({\tilde M}_0)_{21},
~{\tilde{U}}_{41}=({\tilde M}_1)_{21},
~{\tilde{U}}_{13}=({\tilde M}_0)_{12},
~{\tilde{U}}_{23}=({\tilde M}_1)_{12},
~{\tilde{U}}_{33}=({\tilde M}_0)_{22}$,
and ${\tilde{U}}_{43}=({\tilde M}_1)_{22}$
while the elements in the other two columns are yet to be determined.
The second and the fourth columns' elements can be chosen by ensuring $U_{ab}(t)$ to
be unitary. Without loss of generality, we choose $\tilde{U}_{ab}(t)$ as
\begin{widetext}
\begin{eqnarray}
\tilde{U}_{ab}(t)=\left(\begin{array}{cccc}-c_s-r_-s_ce^{i(\phi_0-\phi)}
&-r_+s_se^{-i\phi}&c_ce^{-i\phi_0}-r_-s_se^{-i\phi}
&r_+s_ce^{-i(\phi+\phi_0)}\\ r_+c_ce^{i\phi_0}&-s_ce^{i(\phi_0-\phi)}-r_-c_s&r_+c_s
&-s_se^{-i\phi}+r_-c_ce^{-i\phi_0}\\-s_se^{i\phi}+r_-c_ce^{i\phi_0}&r_+c_s&s_ce^{i(\phi-\phi_0)}+r_-c_s&-r_+c_ce^{-i\phi_0}\\
r_+s_ce^{i(\phi+\phi_0)}&c_ce^{i\phi_0}-r_-s_se^{i\phi}&r_+s_se^{i\phi}&c_s+r_-s_ce^{i(\phi-\phi_0)}\end{array}\right),
\label{tuie}
\end{eqnarray}
\end{widetext}
where $s_s=\sin\frac{\theta}{2}\sin\frac{\theta_0}{2},
~s_c=\sin\frac{\theta}{2}\cos\frac{\theta_0}{2},
~c_s=\cos\frac{\theta}{2}\sin\frac{\theta_0}{2},
~c_c=\cos\frac{\theta}{2}\cos\frac{\theta_0}{2}$
and $r_+=\sqrt{\frac{r+r_0}{1+r_0}},~r_-=\sqrt{\frac{1-r}{1+r_0}}$.

Expression (\ref{tuie}) is only one of the many unitary operators $U_{ab}(t)$
satisfying Eq. (\ref{rhott}). To construct other
equivalent unitary operators,
we need to find other equivalent Kraus operators $M_\mu$.  In terms of the known operators given
by Eq. (\ref{mm}), the equivalent Kraus operators can be written as
\begin{eqnarray}
M_\mu=\sum\limits_{\nu=0,1}W_{\mu\nu}\tilde{M}_\nu V(t),
\label{mwmv}
\end{eqnarray}
where $W_{\mu \nu}$ are the elements of time dependent matrix $W(t)\in SU(2)$, and
$V(t)\in U(2)$ satisfies Eq. (\ref{vr}). The explicit expression of  $V(t)$ is given by
Eq. (\ref{v}) if $r_0\neq 0$ and it can be any arbitrary $2\times 2$ unitary matrix if
$r_0=0$. Eq. (\ref{mwmv}) is the general expression of the Kraus operators realizing
the prescribed evolution of the open system. With the general Kraus operators $M_{\mu}$ , we can
construct the general expression of unitary operators $U_{ab}(t)$.  By using Eq. (\ref{mwmv}),
the equivalent set of the unitary operators can be described by
\begin{eqnarray}
U_{ab}(t)=(I\otimes W(t) )\tilde{U}_{ab}(t)(V(t)\otimes I),
\label{uie}
\end{eqnarray}
Substituting Eq. (\ref{uie}) into Eq. (\ref{h}), one obtains the Hamiltonians
\begin{eqnarray}
H_{ab}(t)&=&i\dot U_{ab}(t)U_{ab}(t)^\dagger  . \label{h3}
\end{eqnarray}
For the prescribed evolution of the open system, there are many
possible choices of Hamiltonians, one of which is given by
$\tilde{H}_{ab}(t)=i{\dot
{\tilde{U}}}_{ab}(t){\tilde{U}}_{ab}(t)^\dagger $ while the
others are obtained by Eq. (\ref{h3}). For a given prescribed
evolution $\rho(t)$, we can immediately write down the
Hamiltonians by substituting the corresponding
$r=r(t),~\theta=\theta (t),~\phi=\phi(t)$ into Eqs. (\ref{tuie}),
(\ref{uie}) and (\ref{h3}). All the Hamiltonians fulfill the
prescribed evolution of the open system. As the matrices $W\in
SU(2)$  and $V(t)=\sum\limits_{k=1,2}e^{i\alpha_k}\rho_k(0)$ if
$r_0\neq 0$, or $V(t)\in U(2)$ if $r_0=0$, there are five or seven
pending parameters to serve any other special requisitions. One
can obtain any required Hamiltonian by choosing properly the
pending parameters.

\section{Example for applications}
In the previous sections, we develop a general procedure of Hamiltonians for
realizing a prescribed evolution of a qubit system, be it closed or open.
In spite of the complicated details, it is still a useful device due to
its generality. As an illustration of its applications, we
consider the measurement of geometric phase as an example.

As we know, when a system evolves along a given path in the project Hilbert space,
the system will acquire the dynamic phase as well as the geometric phase. Geometric phase can be regarded as
the total phase minus dynamic phase. For the mixed state of a qubit system, if the
evolution is defined by Eqs. (\ref{rhot}) and (\ref{rho0}) with $r=r_0$, the geometric
phase acquired by the system is\cite{Sjoqvist,Kuldip}
\begin{eqnarray}
\gamma(\tau)=arg\{\sum\limits_{k=1,2}{\rm
tr}[w_k\rho_k(0)U(\tau)e^{-\int_0^\tau {\rm
tr}[\rho_k(0)U(t)^\dagger \dot U(t)]dt}]\}, \label{g}
\end{eqnarray}
where $w_1=(1+r_0)/2$, $w_2=(1-r_0)/2$, and $\rho_k(0)$ are defined by Eq. (\ref{rho012}).
Mathematically, we may use any one of the unitary operators $U(t)$ given by
Eq. (\ref{u=ur}) to evaluate the geometric phase. All of them result in the same value of  $\gamma(\tau)$. However, the experiment situation for phase measurement
is quite different.
The phase measured in the experiment is usually
the total phase of the system, not the geometric phase. The different choices
of $U(t)$ will lead to different values of the measurement. If one wishes to measure the
geometric phase, one must choose a special unitary operator satisfying the
parallel transport conditions
\begin{eqnarray}
{\rm tr}[\rho_k(0)U(t)^\dagger \dot U(t)]=0,~~(k=1,2).
\label{truu}
\end{eqnarray}
The special unitary operator will render the dynamical phase zero, hence the total phase
equals the geometric phase. Substituting Eqs. (\ref{rho012}) and (\ref{u=ur}) into
Eq. (\ref{truu}), one will find that the parallel transport conditions
constrain the functions $\alpha_1(t)$ and $\alpha_2(t)$ to be \cite{tong}
\begin{eqnarray}
\alpha_1=-\alpha_2=\frac{1}{2}\int_0^t\cos\theta \dot \phi dt.
\label{alpha}
\end{eqnarray}
Substituting Eq. (\ref{alpha}) into Eq. (\ref{hm}), the Hamiltonian is obtained as
\begin{widetext}
\begin{eqnarray}
H(t)=i\dot U(t)U(t)^\dagger =\frac{1}{2} \left(\begin{array}{cc}
\dot \phi\sin^2\theta& (-i\dot \theta-\dot \phi
\sin\theta\cos\theta)e^{-i\phi}
\\ (i\dot \theta-\dot \phi\sin\theta\cos\theta)e^{i\phi} & -\dot \phi\sin^2\theta\end{array}\right),
\label{hm2}
\end{eqnarray}
\end{widetext}
with which the measurement for the total phase results in the geometric phase.
Let us be more specific.
Suppose a qubit system, say an electron,  with initial state $\rho(0)$
($r_0\neq 1$, $\phi_0=0$) given by Eq. (\ref{rho0}), is expected to evolve
along a path with $\theta =\theta_0, ~\phi =\omega t~, ~~t\in [0,~\frac{2\pi}{\omega}]$.
The evolution is a closed circle inside Bloch sphere.
By using expression (\ref{hm2}), the special Hamiltonian can be immediately written down as
\begin{eqnarray}
H(t)= \frac{\omega \sin\theta_0}{2}\left(\begin{array}{cc}
 \sin\theta_0 & -\cos\theta_0 e^{-i\omega t}
\\ -\cos\theta_0 e^{i\omega t} & -\sin\theta_0
\end{array}\right).
\label{hm3}
\end{eqnarray}
The above Hamiltonian ensures the given state to evolve along the prescribed path with zero dynamic
phase. The measurement result of the geometric phase should be
$\gamma=-\tan^{-1}\{r_0\tan [\pi(1-\cos\theta_0)]\}$. This Hamiltonian can be experimentally realized in
physics by applying a time-dependent magnetic field, ${\bf{B}}(t)=
(B_1,~B_2,~B_3)=\omega \sin \theta_0 (-\cos\theta_0 \cos\omega t,
~-\cos\theta_0\sin\omega t,~\sin\theta_0)$, to the electron system, and
$H(t)={\bf \sigma}\cdot {\bf B}(t)/2$.
As the geometric phase has been used as quantum gate, the above discussion may be useful
for quantum computation. As well-known, if an evolution causes nonzero dynamic
phase, one must eliminate it by some means\cite{Falci}-\cite{Ekert}. However, this cancellation is not easy. In our case, given the generic expression for the
Hamiltonian, one can try to select those Hamiltonians for which the dynamic phase is zero.
Recall that Eqs. (\ref{hm2}) and (\ref{hm3}) give no dynamical phase for any path of the state,
be it closed or open. It is also interesting to consider non-cyclic evolutions instead of
cyclic closed circuit. In this context, we would like to mention a recent paper\cite{Zhu},
where it is argued that some systems may obey the same fault-tolerance properties like those of
the geometric phase also for nonvanishing dynamical phase.

The above example provides a concrete application of our formalism 
in the form of the geometric phase of mixed state under
unitary evolution. Similarly, we can also consider examples
of mixed state under nonunitary evolutions. For instance, suppose we need to prepare a
qubit system whose Bloch vector ${\bf{r}}$ is expected to evolve along a prescribed
path inside  the Bloch sphere, say, an ellipse with
$r_x^2+4r_y^2=1,~r_z=0$.
We can realize this evolution through the combined system of the qubit and
an ancillary qubit. The unitary operator ${\tilde{U}_{ab}}(t)$ can be easily
calculated by substituting $r=(\cos^2\omega t +4\sin^2\omega t)^{\frac{-1}{2}},
~\theta=\frac{\pi}{2},~\phi=\omega t,$ and $r_0=1,~\theta_0=\frac{\pi}{2},~\phi_0=0$
into Eq. (\ref{tuie}). With $ \tilde{U}_{ab}(t)$, the general unitary operators
$U_{ab}(t)$ and the general Hamiltonian $H_{ab}(t)$ can be obtained by Eq. (\ref{uie})
and Eq. (\ref{h3}) respectively. Each Hamiltonian $H(t)$ obtained in this way
satisfies the same evolution with the appropriate reduced matrix $\rho(t)$
along the given ellipse. If there are other physical
constraints, one can then choose a preferred Hamiltonian by determining the suitable
parameters in the Hamiltonians. As the Hamiltonians involve complex $4\times 4$ matrices
again, we will not describe the example in detail in the present paper.
Instead, we prefer to show an explicit form of the Hamiltonians for a simple
example, the pure shrinking of the Bloch vector along the polar axis. In this case,
by putting $\theta=\theta_0=0,~r_0=1$ in Eq. (\ref{tuie}),
the Hamiltonian $\tilde{H}_{ab}(t)$ is obtained as
\begin{eqnarray}
\tilde{H}_{ab}(t)=i{\dot
{\tilde{U}}}_{ab}(t){\tilde{U}}_{ab}(t)^\dagger =\frac{\dot
r}{4\sqrt{1-r^2}}[\sigma_x\otimes \sigma_y-\sigma_y\otimes
\sigma_x], \label{simple}
\end{eqnarray}
where $\sigma_x,~ \sigma_y$ are Pauli matrixes,
and $r~ (0\leq r\leq 1)$ is an arbitrary prescribed time dependent function
with $\dot r\leq 0$. $\tilde{H}_{ab}(t)$ will render the Bloch vector of the state shrink
from the pole $(1,0,0)$ towards the center of the Bloch sphere with the prescribed speed
$|\dot r|$. Other Hamiltonians equivalent to $\tilde{H}_{ab}(t)$ can be obtained by using
Eqs. (\ref{uie})  and (\ref{h3}).

\section{Discussions and Conclusions}
The inverse problem for the evolution of a qubit is investigated in this paper.
We proposed a general formalism to establish the Hamiltonian for a qubit system
that is required to evolve along a particular path. Both unitary and nonunitary
evolutions of mixed states of a qubit system are discussed.

For a closed qubit system, its evolution is unitary. The general form of Hamiltonians
for realizing a prescribed evolution is explicitly given by Eq. (\ref{hm}).
One can directly write down the Hamiltonians by substituting $\theta=\theta(t)$ and $\phi=\phi(t)$ into Eq. (\ref{hm}).
The explicit expression for the Hamiltonian contains two arbitrary time dependent parameters.
Regardless of the values of the parameters, all Hamiltonians give rise to
the same evolution. One may choose appropriate parameters to satisfy the physical constraints.

For an open qubit system, its evolution is generally nonunitary. However the system with its
environment continues to evolve as a closed system. We therefore investigate the open
system by considering the closed system constructed by adding an ancillary qubit to the open
system. The evolution of the open system is expressed by the reduced density matrix
given by Eq. (\ref{rhott}). If the desired evolution is as given by Eqs. (\ref{rhot}) and (\ref{rho0}),
the unitary operators $U_{ab}(t)$ and the Hamiltonians $H_{ab}(t)$ can be calculated
by Eqs. (\ref{uie}) and (\ref{h3}). The expression of the Hamiltonians
contain five or seven arbitrary parameters.  All Hamiltonians of the given form
cause the qubit to evolve along the desired path  while they really act 
on the combined system. One may choose the suitable parameters for the Hamiltonians
to meet any physical constraints, just like the unitary case.

As an example for illustrating the applications of our result, we discuss
the parallel transport conditions on geometric phase measurement. A general expression of
the Hamiltonians satisfying the conditions is given by Eq. (\ref{hm2}). With the kind of
Hamiltonian, the value of the measurement on total phase results in the geometric phase.

The study in present paper focuses on the qubit system. However, the approach can be generalized
to higher dimensional systems. Some main formulae are similar. In fact, for unitary evolution of
$N$-dimension system,  Eqs. (\ref{v})-(\ref{h=iuu}) are still valid with $k=1,2,...,N$.
For nonunitary evolution of $N$-dimension open system, Eqs. (\ref{uie}) and (\ref{h3}) still hold.
The difference for an $N$-dimension system from the qubit system appears
in $\tilde{U}(t)$ given by Eq. (\ref{tu}) or $\tilde{U}_{ab}(t)$ given by Eq. (\ref{tuie}),
which will be replaced by some new matrices corresponding to an $N$-dimensional system.

\section*{Acknowledgments}
The work was supported by NUS Research Grant No. R-144-000-071-305.
JLC acknowledges financial support from Singapore Millennium Foundation.


\begin{thebibliography}{99}
\bibitem{tong} The discussion is for mixed state, $r\neq 1$. For pure state, $r=1$, the
constraint is that $\alpha_1 =\frac{1}{2}\int_0^t\cos\theta~ \dot \phi dt$~while~$\alpha_2 $ is arbitrary.
\end{thebibliography}
\end{document}